\documentclass[twoside,twocolumn,9pt]{article}
\usepackage{extsizes}
\usepackage[super,sort&compress,comma]{natbib} 
\usepackage[version=3]{mhchem}
\usepackage[left=1.5cm, right=1.5cm, top=1.785cm, bottom=2.0cm]{geometry}
\usepackage{balance}
\usepackage{widetext}
\usepackage{times,mathptmx}
\usepackage{sectsty}
\usepackage{graphicx} 
\usepackage{lastpage}
\usepackage[format=plain,justification=raggedright,singlelinecheck=false,font={stretch=1.125,small,sf},labelfont=bf,labelsep=space]{caption}
\usepackage{float}
\usepackage{fancyhdr}
\usepackage{fnpos}
\usepackage[english]{babel}
\usepackage{array}
\usepackage{droidsans}
\usepackage{charter}
\usepackage[T1]{fontenc}
\usepackage[usenames,dvipsnames]{xcolor}
\usepackage{setspace}
\usepackage{draftwatermark}
\SetWatermarkText{Author's Postprint}
\SetWatermarkScale{0.5}
\SetWatermarkAngle{45}
\SetWatermarkColor[gray]{0.8}
\usepackage[compact]{titlesec}

\graphicspath{{Figures/}}

\usepackage{epstopdf}

\definecolor{cream}{RGB}{222,217,201}

\begin{document}

\pagestyle{fancy}
\thispagestyle{plain}
\fancypagestyle{plain}{

\fancyhead[C]{\includegraphics[width=18.5cm]{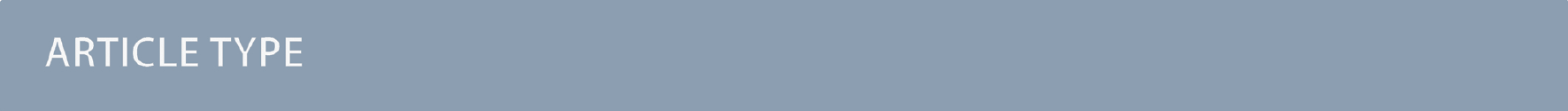}}
\fancyhead[L]{\hspace{0cm}\vspace{1.5cm}\includegraphics[height=30pt]{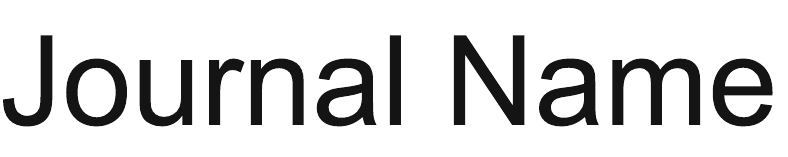}}
\fancyhead[R]{\hspace{0cm}\vspace{1.7cm}\includegraphics[height=55pt]{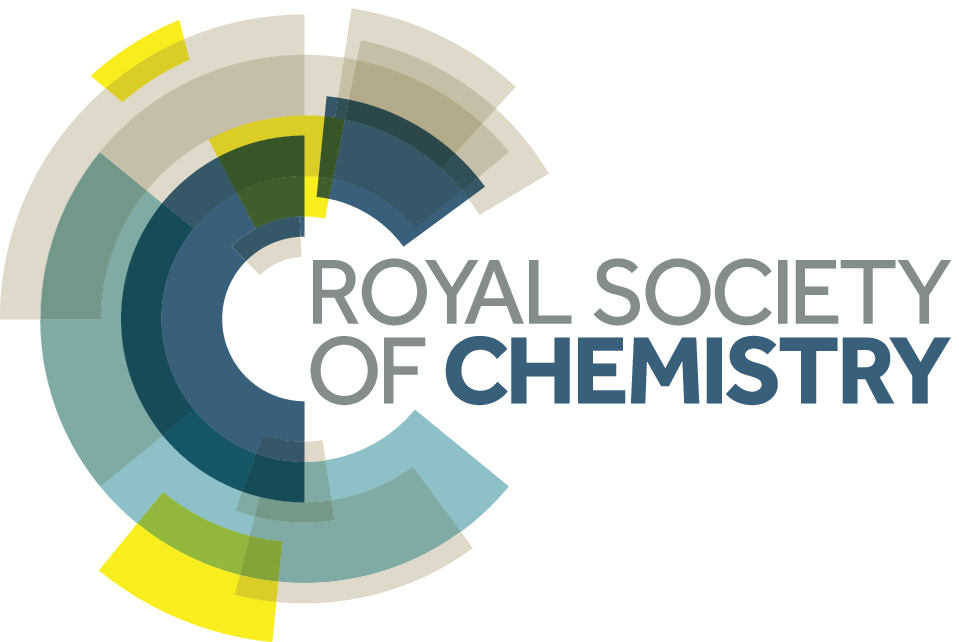}}
\renewcommand{\headrulewidth}{0pt}
}

\makeFNbottom
\makeatletter
\renewcommand\LARGE{\@setfontsize\LARGE{15pt}{17}}
\renewcommand\Large{\@setfontsize\Large{12pt}{14}}
\renewcommand\large{\@setfontsize\large{10pt}{12}}
\renewcommand\footnotesize{\@setfontsize\footnotesize{7pt}{10}}
\makeatother

\renewcommand{\thefootnote}{\fnsymbol{footnote}}
\renewcommand\footnoterule{\vspace*{1pt}%
\color{cream}\hrule width 3.5in height 0.4pt \color{black}\vspace*{5pt}} 
\setcounter{secnumdepth}{5}

\makeatletter 
\renewcommand\@biblabel[1]{#1}            
\renewcommand\@makefntext[1]%
{\noindent\makebox[0pt][r]{\@thefnmark\,}#1}
\makeatother 
\renewcommand{\figurename}{\small{Fig.}~}
\sectionfont{\sffamily\Large}
\subsectionfont{\normalsize}
\subsubsectionfont{\bf}
\setstretch{1.125} 
\setlength{\skip\footins}{0.8cm}
\setlength{\footnotesep}{0.25cm}
\setlength{\jot}{10pt}
\titlespacing*{\section}{0pt}{4pt}{4pt}
\titlespacing*{\subsection}{0pt}{15pt}{1pt}

\fancyfoot{}
\fancyfoot[LO,RE]{\vspace{-7.1pt}\includegraphics[height=9pt]{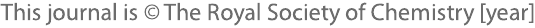}}
\fancyfoot[CO]{\vspace{-7.1pt}\hspace{13.2cm}\includegraphics{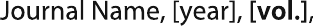}}
\fancyfoot[CE]{\vspace{-7.2pt}\hspace{-14.2cm}\includegraphics{head_foot/RF}}
\fancyfoot[RO]{\footnotesize{\sffamily{1--\pageref{LastPage} ~\textbar  \hspace{2pt}\thepage}}}
\fancyfoot[LE]{\footnotesize{\sffamily{\thepage~\textbar\hspace{3.45cm} 1--\pageref{LastPage}}}}
\fancyhead{}
\renewcommand{\headrulewidth}{0pt} 
\renewcommand{\footrulewidth}{0pt}
\setlength{\arrayrulewidth}{1pt}
\setlength{\columnsep}{6.5mm}
\setlength\bibsep{1pt}

\makeatletter 
\newlength{\figrulesep} 
\setlength{\figrulesep}{0.5\textfloatsep} 

\newcommand{\topfigrule}{\vspace*{-1pt}%
\noindent{\color{cream}\rule[-\figrulesep]{\columnwidth}{1.5pt}} }

\newcommand{\botfigrule}{\vspace*{-2pt}%
\noindent{\color{cream}\rule[\figrulesep]{\columnwidth}{1.5pt}} }

\newcommand{\dblfigrule}{\vspace*{-1pt}%
\noindent{\color{cream}\rule[-\figrulesep]{\textwidth}{1.5pt}} }

\makeatother

\twocolumn[
  \begin{@twocolumnfalse}
\vspace{3cm}
\sffamily
\begin{tabular}{m{4.5cm} p{13.5cm} }

\includegraphics{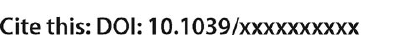} & \noindent\LARGE{\textbf{Two-dimensional pattern formation in ionic liquids confined between graphene walls}} \\
\vspace{0.3cm} & \vspace{0.3cm} \\

 & \noindent\large{Hadri\'an Montes-Campos,\textit{$^{a}$} Jos\'e Manuel Otero-Mato,\textit{$^{a}$} Trinidad M\'{e}ndez-Morales,\textit{$^{a}$} Oscar Cabeza,\textit{$^{b}$} Luis J. Gallego,\textit{$^{a}$} Alina Ciach,\textit{$^{c}$} and Luis M. Varela$^{\ast}$\textit{$^{a}$}} \\

\includegraphics{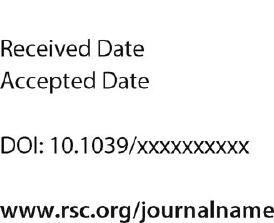} & \noindent\normalsize{We perform molecular dynamics simulations of ionic liquids confined between graphene walls under a large variety of conditions (pure ionic liquids, mixtures with water and alcohols, mixtures with lithium salts and defective graphene walls). Our results show that the formation of striped and hexagonal patterns in the Stern layer can be considered as a general feature of ionic liquids at electrochemical interfaces, the transition between patterns being controlled by the net balance of charge in the innermost layer of adsorbed molecules. This explains previously reported experimental and computational results and, for the first time, why these pattern changes are triggered by any perturbation of the charge density at the innermost layer of the electric double layer (voltage and composition changes, and vacancies at the electrode walls, among others), which may help tuning electrode-ionic liquid interfaces. Using Monte Carlo simulations we show that such structures can be reproduced by a simple two-dimensional lattice model with only nearest-neighbour interactions, governed by highly screened ionic interactions and short-range and excluded volume interactions. We also show that the results of our simulations are consistent with those inferred from the Landau-Brazovskii theory of pattern formation in self-assembling systems. The presence of these patterns at the ionic liquid graphene-electrode interfaces may have a strong impact on the process of ionic transfer from the bulk mixtures to the electrodes, on the differential capacitance of the electrode-electrolyte double layer or on the rates of redox reactions at the electrodes, among other physicochemical properties, and is therefore an effect of great technological interest. }\\ 

\end{tabular}

 \end{@twocolumnfalse} \vspace{0.6cm}

  ]

\renewcommand*\rmdefault{bch}\normalfont\upshape
\rmfamily
\section*{}
\vspace{-1cm}


\footnotetext{\textit{$^{a}$~Grupo de Nanomateriales, Fot\'onica y Materia Blanda. Departamentos de F\'isica de Part\'iculas y F\'isica Aplicada, Facultade de F\'isica, Universidade de Santiago de Compostela, Campus Vida s/n, E-15782 Santiago de Compostela, Spain.}}
\footnotetext{\textit{$^{b}$~Grupo Mesturas, Universidade da Coru\~na, Departamento de F\'isica e Ciencias da Terra, Facultade de Ciencias, Campus da
Zapateira, 15071 A Coru\~na, Spain. }}
\footnotetext{\textit{$^{c}$~Institute of Physical Chemistry, Polish
  Academy of Sciences, Kasprzaka 44/52, 01-224 Warszawa, Poland. }}

\footnotetext{\textit{$^{\ast}$~Corresponding author, to which correspondence should be addressed. E-mail: }\mbox{luismiguel.varela@usc.es}}

\section{Introduction}
Ionic liquids (ILs) have attracted, and continue to attract, considerable
research efforts both for theoretical reasons (the desire to understand the
peculiar properties of these complex systems) and for their many potential
applications, including as solvents for synthesis and catalysis, thermal fluids
for sensible heat storage in thermosolar power systems, lubricants and,
prominently, as advanced electrolyte materials for fuel cells, batteries or
supercapacitors, among others (see, e.g., refs. \citenum{Welton-1999,
Earle-2000, Rogers-2003, Galinski-2006, Parvulescu-2007, Plechkova-2008,
Greaves-2008, Weingartner-2008, Armand-2009}). Beyond their behaviour at bulk
level, the properties of ILs at interfaces are specially relevant since it is at
the interface where redox reactions usually take place. Until now, most studies
in this area have been focused on the oscillations of charge in the direction
perpendicular to the interface and their effect in the adsorption of cosolvents
at the electrode (see the recent review of Fedorov and Kornyshev
\cite{Fedorov-2014} and references therein). However, despite it probably exerts
a profound influence on charging dynamics, \cite{dudka-2016} the lateral
structure of ILs nanoconfined in the adlayer has received much less attention.
Merlet \emph{et al}.\cite{Merlet-2014} and Rotenberg and
Salanne\cite{rotenberg-2015} reported molecular dynamics (MD) studies of the
interface between buthylmethylimidazolium hexafluorophosphate and graphite and
Au(111) electrodes at constant potential, showing the occurrence of in-plane
structures and fluctuations beyond mean-field in the first adsorbed layer of the
IL. The authors analysed voltage driven changes in the lateral charge
distribution that induce transitions form ordered states into disordered
nonstoichiometric states, leading to peaks in the capacitance. These
configurations were similar to those experimentally detected by Liu \emph{et
al.}\cite{Liu-2006} and more recently by Ebeling \emph{et
al}.\cite{Ebeling-2016} Using atomic force microscopy, the latter authors
investigated the lateral structure of the innermost layer of the IL-solid
interface and the associated three-dimensional (3D) structure induced by
molecule-substrate interactions, revealing quasi (4$\times$4)R0$^{\circ}$
overlayer. On the other hand, Dudka \emph{et al.}\cite{dudka-2016} have used a
statistical theory based on bipartite lattices and a 3D off-lattice Monte Carlo
(MC) simulation of the structure and phase behaviour of superionic liquids in
nonpolarized nanoconfinement, showing the existence of crystal-like (ordered)
and homogeneous (disordered) phases separated by first or second order phase
transitions. {Also on the theoretical side, Limmer\cite{Limmer-2015}
reported an effective field theory predicting a fluctuation-induced, first-order
interfacial transition associated with spontaneous ordering of the ions in the
layer nearest the electrode, which explains the ``anomalous'' behaviour of the
differential capacitance of the IL-based electric double layer. The author
showed that two opposing interactions are behind this first-order interfacial
transition: short-range interactions between like species leading to the phase
separation, and long-range electrostatic interactions between ionized species,
leading to the neighbourhood of opposite charges. The combined action of these
two interactions induces long-ranged order (nonvanishing charge for vanishing
potential) in a slab parallel to the electrode whose width equals the bare
correlation length. However, neither the theoretical analysis nor the MC
simulations  provide any direct evidence of the lateral patterns in the
near-electrode layer.} On the other hand, by means of MD simulations Docampo
\emph{et al.}\cite{Docampo-2016} have shown that the lateral structure of the
first layer of anions close to a graphene surface undergoes a transition from
ordered stripes to an ordered phase with hexagonal symmetry upon the addition of
water. Hence, the authors pioneeringly showed the existence of concentration
driven structural transitions in the adlayer structure, which were recently
complemented by Gomez-Gonzalez \emph{et al.},\cite{Gomez-Gonzalez-2017} who
reported the occurrence of these transitions also in mixtures of ILs with
monovalent and divalent salts at the graphene interface.

These structural changes at IL-graphene interfaces seem to be a quite general emergent pattern of nanoconfined ILs that could have
a strong impact on the physical and chemical
properties, such as the capacitance and the mechanical response, which could be of great importance for electrochemical devices, voltage- and composition-dependent lubricants and actuators. So, although the interface structure may depend on the nature of the electrode and the IL-electrode interaction, it is important to know if there are some 
general features, for which we will consider the particularly relevant IL-graphene interface, and, if this is so, to give a proper theoretical interpretation of that behaviour. Moreover, the elucidation of the actual role played by the surface on these patterns is another critical issue still to be appropriately addressed. 


 
In this work we use MD simulations to show that
the formation of ordered patterns at IL interfaces is caused not only
by the addition of water or by changes in the electrode potential, but also by changes in the structure of the electrodes or
by the addition of other cosolvents such as alcohols or inorganic
salts; indeed, by any other perturbation which alters the balance of charge in the adlayer. Additionally, by means of NVT MC simulations we show
that the main features of such structural behaviour can be reproduced
by a simple two-dimensional (2D) lattice model with only
nearest-neighbour interactions and report a schematic phase diagram.
Moreover, we propose a theoretical explanation of this
behaviour using the Landau-Brazovskii phenomenological theory,
\cite{brazovskii:75:0} which has been previously used to explain
pattern formation in self-assembling systems.
\cite{leibler:80:0,fredrickson:87:0,seul:95:0,gompper:94:3,ciach:01:2}

The essential technical details of the computational methods used are
sketched in the next section, our results are presented and discussed
in the section 3, and finally we summarize our main
conclusions.

\section{Computational Details}
\label{sec:comp_details}

MD simulations were carried out using Gromacs 5.0\cite{Gromacs} in a
NVT ensemble at room temperature (298 $K$), which was kept constant using a V-rescale\cite{bussi2007canonical} thermostat. In
order to avoid the system being trapped in a local metastable
configuration, it was annealed up to 600 $K$ for 1 ns before
the production run. The simulation system was formed by
\ce{[BMIM][BF4]} (1-butyl-3-methylimidazolium tetrafluoroborate) with
different cosolvents (water, methanol, ethanol and lithium
tetrafluoroborate) between two graphene electrodes. The number of IL pairs
was set to 900 for the pure IL and its mixtures with
\ce{LiBF4}. For the other cases, it was set to 950. These numbers were
chosen so the number of dissolved molecules was at least 50 in order to have reliable enough statistical averages of their magnitudes. The exact number of molecules used for each system is shown in Table \ref{tab:molecules}.

\begin{table}[h]
  \centering
    \caption{Number of molecules used in the reported MD simulations.}
  \begin{tabular*}{8.5cm}{ccc}
    \hline
    System& IL molecules & Dissolved molecules \\\hline
    Pure IL & 900 & - \\
    5\% Water-IL & 950 & 50 \\
    5\% methanol-IL & 950 & 50 \\
    5\% ethanol-IL & 950 & 50 \\
    10\% \ce{LiBF4}-IL + \\  5\% vacancies & 900 & 100 \\
    10\% \ce{LiBF4}-IL-IL+ \\  8\% vacancies & 900 & 100 \\
    \hline
  \end{tabular*}

  \label{tab:molecules}
\end{table}

All interactions were represented by means of the OPLS-AA force
field.\cite{jorgensen1996development} The electrodes were modeled as a
rigid graphene sheet with a charge of $\pm 1 ~e~\textrm{nm}^{-2}$. The pristine
graphene sheets were generated using the Visual Molecular Dynamics
(VMD) software.\cite{humphrey1996vmd} The vacancy defects on the
electrodes were introduced by randomly removing carbon atoms from the
pristine graphene sheet and distributing their charge among their
first neighbours. Furthermore, an additional constraint in order to
avoid the appearance of isolated atoms was used: the first and second
neighbours of a removed carbon atom were excluded from subsequent removals.

The initial configurations were generated using Packmol\cite{packmol}
with a distance between the graphene electrodes ca. 10 nm. Their lateral sizes were then chosen in order to reproduce the average bulk density of the corresponding mixture in the center of the box. After that, the systems were annealed up to 600 K during 1 ns, equilibrated at room temperature for 10 ns and, finally, production runs of 5 ns with steps of 2 fs were performed.

On the other hand, MC simulations were performed for a model Coulomb
system composed of anions and cations distributed in a triangular
lattice of 400 nodes in the NVT ensemble using the standard Metropolis
algorithm. In each simulation step, two nearest-neighbour ions are allowed to swap positions, the move being accepted according to the normal Metropolis algorithm. The simulations consisted of 250000 MC stabilization loops (where loop means a number of steps equal to the number of nodes) and a production run of 250000 additional loops. The interaction energy was
calculated using a lattice model similar to that reported by Dudka\emph{ et
al.}, \cite{dudka-2016} as described in detail in Sec. 3.

\section{Results and Discussion}

In order to address the challenges mentioned in the Introduction, we
first performed MD simulations to analyse the adlayer structure of different mixtures of \ce{[BMIM][BF4]} confined between both pristine and defective (with vacancies) graphene electrodes. Specifically, we investigated the behaviour of the pure IL and its mixtures with water, ethanol or methanol confined between two graphene sheets with pristine structures. Moreover, we investigated both the pure IL and its mixtures with a Li salt (\ce{LiBF4}) when one of the graphene sheets has different percentages of vacancy defects, in order to test the influence of the reduction of the surface symmetry. We
note that the presence of defects in graphene-like materials are
sometimes unavoidable, but can also be intentionally added for
specific practical purposes (see, e.g., refs. \citenum{araujo-2012} and
\citenum{terrones-2012}).
[BMIM]\ce{[BF4]} was chosen because its cation is among the most
commonly used and its anion has tetrahedral symmetry, which makes it
specially suitable for structural studies. The vacancies were randomly
distributed in the surface and the partial charge of the removed atoms was homogeneously distributed among their closest neighbours.


\begin{figure}
  \begin{minipage}{4.12cm}
    \includegraphics[width=\linewidth]{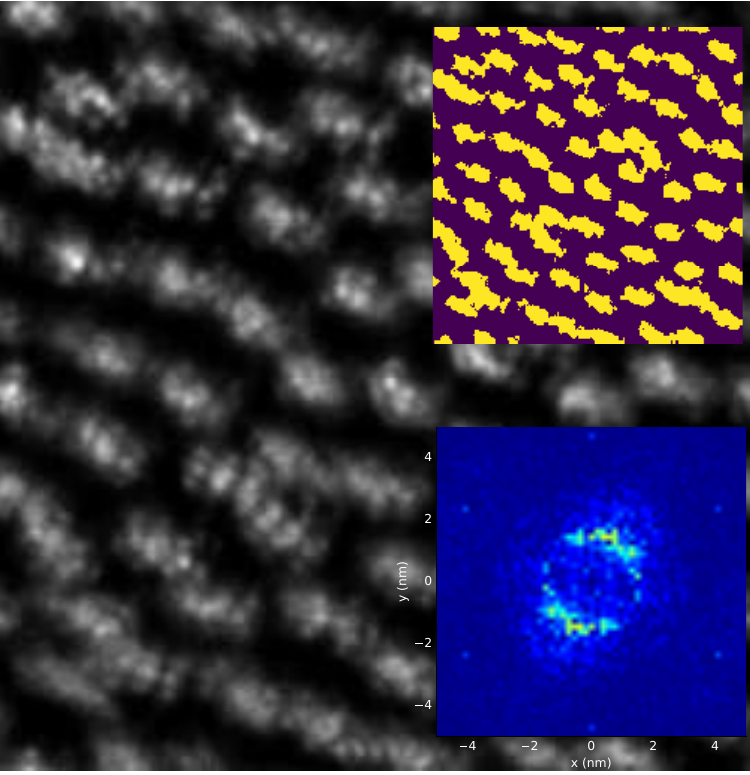}
  \end{minipage}
  \begin{minipage}{4.12cm}
    \includegraphics[width=\linewidth]{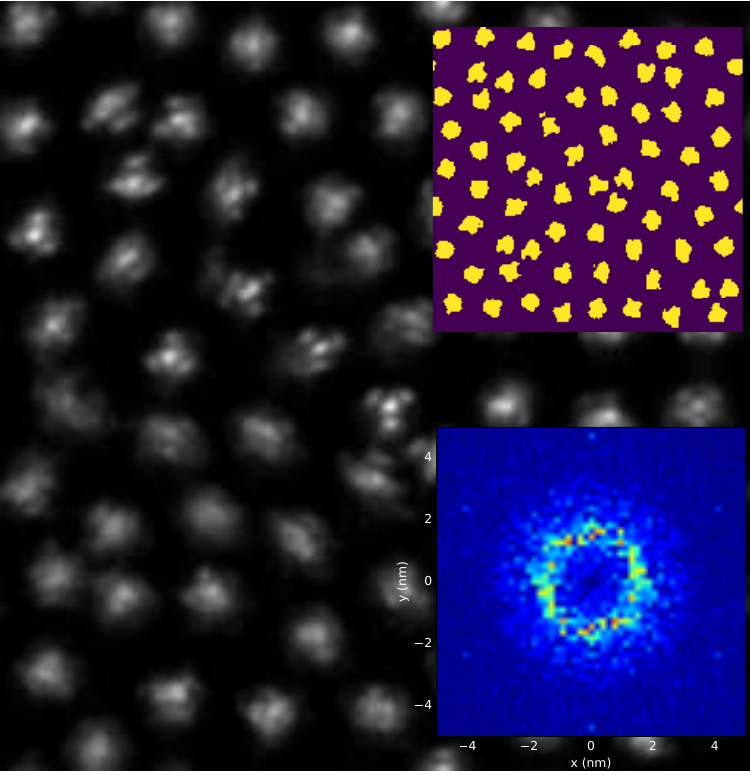}
  \end{minipage}
  \caption{Gray scale representation of the number density in the layer of thickness 5 \AA~ of a mixture of \ce{[BMIM][BF4]} with a 10\% of \ce{LiBF4} closest to a graphene cathode with 5\% (left) and 8\%
    (right) of vacancy defects. The insets correspond to the results
    of applying Otsu's algorithm (top) and the fast 2D Fourier transform (bottom). The left image shows an example of striped phase and the right one of an hexagonal phase.}
    \label{fig:example}
\end{figure}

In order to analyse the lateral structure of the mixtures in the adlayers, we
computed the surface number density of the species in the mixtures in the 5 \AA~
slab closest to the electrode in all cases, which corresponds to the first
minimum of the density for mixtures with water,\cite{Docampo-2016} and, in
general, is a good approximation for most mixtures. Illustrative results for a
mixture of \ce{[BMIM][BF4]} and \ce{[Li][BF4]} next to the graphene cathode with
different percentages of vacancy defects are displayed in Fig.
\ref{fig:example}. The MD results in gray show that the systems have an
amorphous structure, so the Fourier transform (which gives the in-plane
structure factor) is not very informative, at least for the electrode sizes
achievable with our fully atomistic MD simulations. However, a more clear
description of the patterns can be obtained by calculating the Minkowski
functionals \cite{mecke-1996} after treating the MD images with Otsu's
algorithm, \cite{otsu-1975} which uses Fisher's discriminant analysis to
separate the image into two populations. {It is one of the most frequently
used  methods to reduce a gray level image to a binary one by means of automatic
clustering-based image thresholding. The core of the algorithm is finding the
optimal assignment in a binary image of the pixels of a gray scale one. For
that, the image is assumed to contain only two different populations of pixels,
foreground and background, with a bi-modal histogram, so a threshold must be
defined in order to univocally assign pixels of the original image to one
population or the other. Afterwards, the threshold that minimizes the
intra-class variance, i.e., the variance of the distribution of pixels in each
side of the threshold, is to be found. This calculation can be equivalently and
more efficiently done maximizing the inter-class variance, since the sum of
pairwise squared distances is constant.} The results of these calculations
clearly show the presence of striped and hexagonal phases, similar to those
reported in ref. \citenum{rotenberg-2015}, depending of the vacancy defect
percentage. Similar patterns were obtained in all the studied cases as shown in
Table \ref{tab:phi}. 


An appropriate magnitude to characterise the two phases mentioned above is the
surface fraction occupied by the anions, $\eta$, which takes values close to 0.3
for the striped phase and to 0.2 for the hexagonal phase. {The values of $\eta$
for the different investigated systems were obtained using the previously
described Otsu's algorithm to automatically calculate the surface area occupied
by the anions, which was later divided by the corresponding surface area of the
graphene wall; the} results are shown in Table \ref{tab:phi}. It must be
mentioned that the mixture with water is an intermediate state between both
structures (molten lamella\cite{Pekalski-2014,almarza-2014}), i.e., it
corresponds with a limiting state between the striped and hexagonal patterns.

\begin{figure}
  \includegraphics[width=8.25cm]{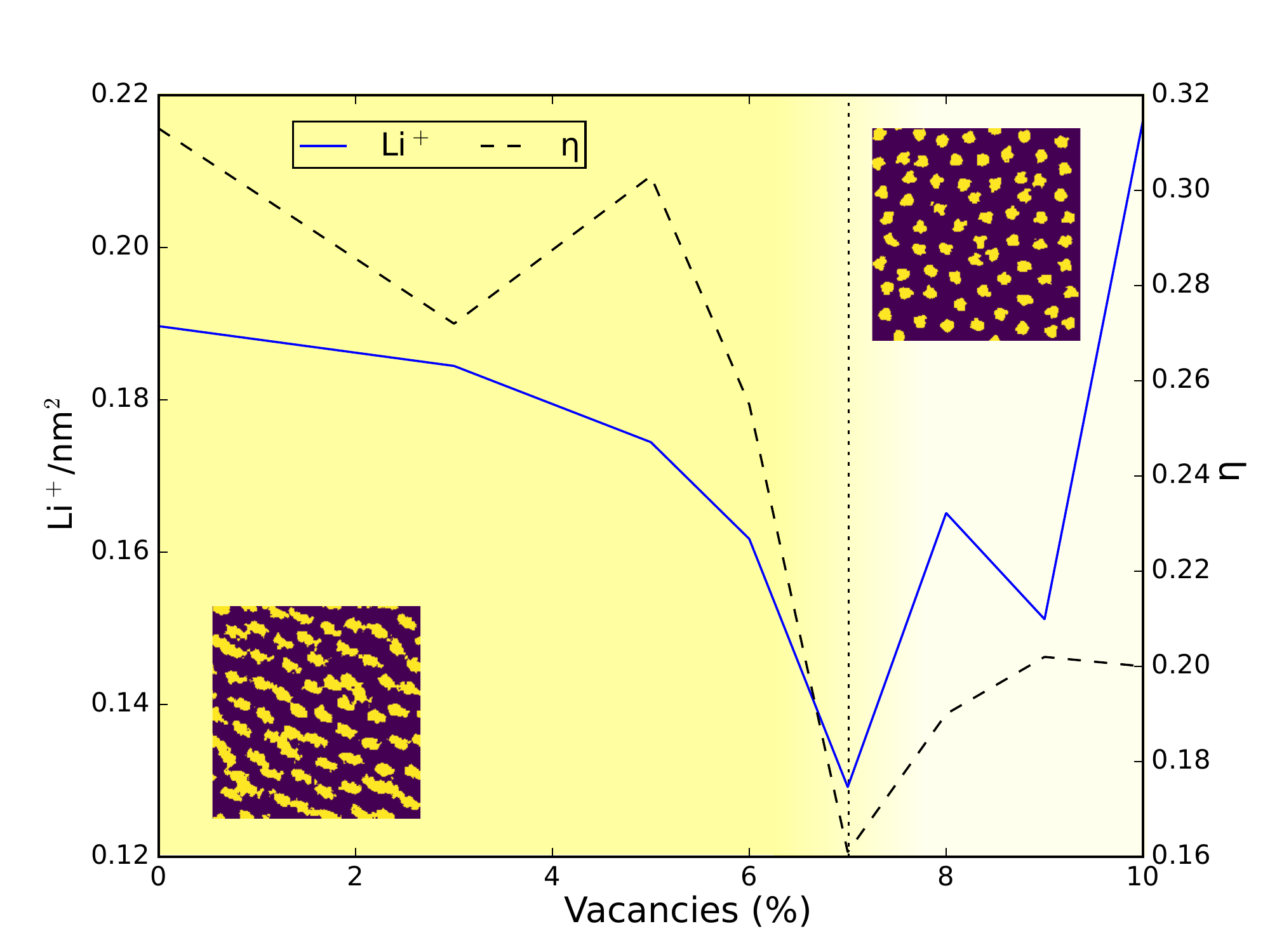}
  \caption{Concentration of adsorbed Li$^{+}$ species {in the 10\% [BMIM][BF$_4$] + LiBF$_4$ mixture} as a function of the fraction of vacancies in the positively charged wall compared with the mean value of the surface fraction covered by the anions, $\eta$. The vertical dotted line corresponds to the percentage of defects that is limiting between the striped and hexagonal adlayer patterns.}
  \label{fig:adsorption}
\end{figure}

In order to see if these structural transitions have any actual influence on other physically relevant parameters, we analyzed the presence of
\ce{Li} in the adlayer for mixtures of \ce{[BMIM][BF4]} and
\ce{LiBF4} as a function of the number of vacancies, and the results
are shown in
Fig. \ref{fig:adsorption}. There we observe that surface
inhomogeneities can change the amount of cosolvent adsorbed at the
positively charged wall: a clear change in the number of adsorbed \ce{Li+} species in the mixture \ce{[BMIM][BF4]} + \ce{LiBF4} is registered at 7\% of vacancies in the graphene cathode, which is the concentration for which the transition from the striped to the hexagonal phase takes place. It should also be pointed out that, in general, the effect of these patterns does not end in the first layer of the IL, but it should determine the 3D structure of the
electric double layer of the electrode, as pointed out by Ebeling \emph{et al.}\cite{Ebeling-2016}



\begin{table}[h]

  \caption{Values of the average fraction of surface covered by anions
    ($\eta$) in mixtures of \ce{[BMIM]}\ce{[BF4]}. The vacancy
    percentages correspond to those in the positively charged graphene
   wall.}
  \label{tab:phi}
  \begin{center}
 \begin{tabular*}{0.65\columnwidth}{ccc}
  \hline\hline
    Mixture &$\eta$ & Phase\\ \hline
    Pure & 0.34 & Stripes\\
   5\% water & 0.36 & Molten Lamella\\
    5\% methanol & 0.29 & Stripes\\
    5\% ethanol & 0.16 & Hexagons\\
    5\% vacancies & 0.19 & Hexagons\\
        8\% vacancies & 0.17 & Hexagons\\
    10\% Li salt +\\ 5\% vacancies & 0.30 & Stripes\\
    10\% Li salt +\\ 8\% vacancies & 0.19 & Hexagons\\
    \hline\hline
  \end{tabular*}
  \end{center}
\end{table}

For further exploring, under the simplest possible conditions, the
mechanism of pattern formation at IL interfaces and to build the phase
diagram of the system, we performed NVT MC simulations using
a 2D lattice model similar to that reported by Dudka \emph{et al.}
\cite{dudka-2016}, but using only two species (anions and cations)
since we restricted ourselves to the simplest case of pure IL
with nearest-neighbour interactions only. This latter assumption is realistic
for ILs since the high charge density in the studied nanoconfined systems gives
rise to highly screened Coulomb interactions and, hence, to very short screening
lengths, so, in practice, only nearest-neighbour interactions need to be taken
into account.\cite{Ebeling-2016} This kind of analytically tractable simplified
lattice model has been quite commonly used for qualitative (and even
quantitative) describing dense ILs, one-dimensional Coulomb systems as single
file pores and ultrathin slits (see ref. \citenum{Ebeling-2016} and references
cited therein). {Similar lattice calculations were performed by
Limmer\cite{Limmer-2015} using a 3D charge frustrated Ising model.} Of course,
repulsive anion-anion and cation-cation interactions ($I_{nn}>0$ and $I_{pp}>0$)
and attractive anion-cation interactions ($I_{np}<0$) define the physical
region, being the total energy of the system
\begin{equation}
  \label{eq:total_energy_1}
  E/k_BT=N_{pp}I_{pp}+N_{nn}I_{nn}+N_{np}I_{np},
\end{equation}
where $N_{nn}$, $N_{pp}$ and $N_{np}$ are the number of anion-anion,
cation-cation and anion-cation pairs, respectively, and $k_B$ is the
Boltzmann constant. The species are considered to be distributed at
the nodes of a triangular lattice with periodic boundary conditions.
This lattice was chosen since its symmetry is that of the adsorption
sites (hole sites) of the graphene sheet. However, as mentioned above, the actual in-layer structures observed for pristine graphene and for graphene with vacancies suggest that, expectedly,  the results are quite independent of the actual electrode's lattice. In each simulation step, one node is chosen randomly, as it is one of its first neighbours. Then, an attempt to swap the nodes is accepted according to the conventional Metropolis algorithm.

\begin{figure}[h!]
  \includegraphics[width=8.25cm]{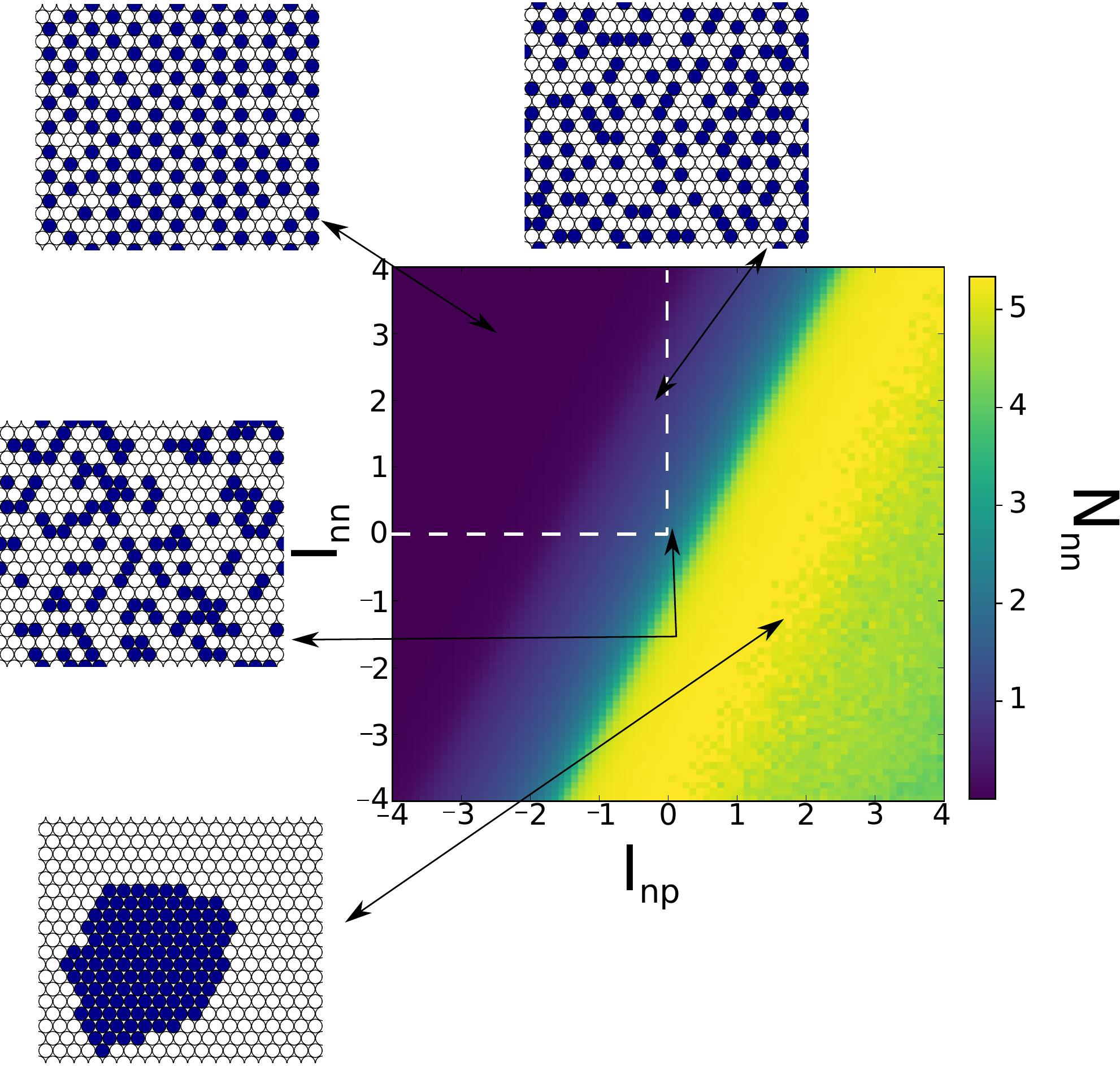}
  \caption{Phase diagram for the lattice system defined by eq.
    \ref{eq:total_energy_1} at a fraction of anions of 0.3 ($I_{nn}$
    and $I_{np}$ in units of $k_BT$). Snapshots of the system are
    shown for different values of $N_{nn}$. Dashed lines indicate the
    borders of the physically meaningful region.}
  \label{fig:phase_diagram}
\end{figure}

\begin{figure}
  \includegraphics[width=8.25cm]{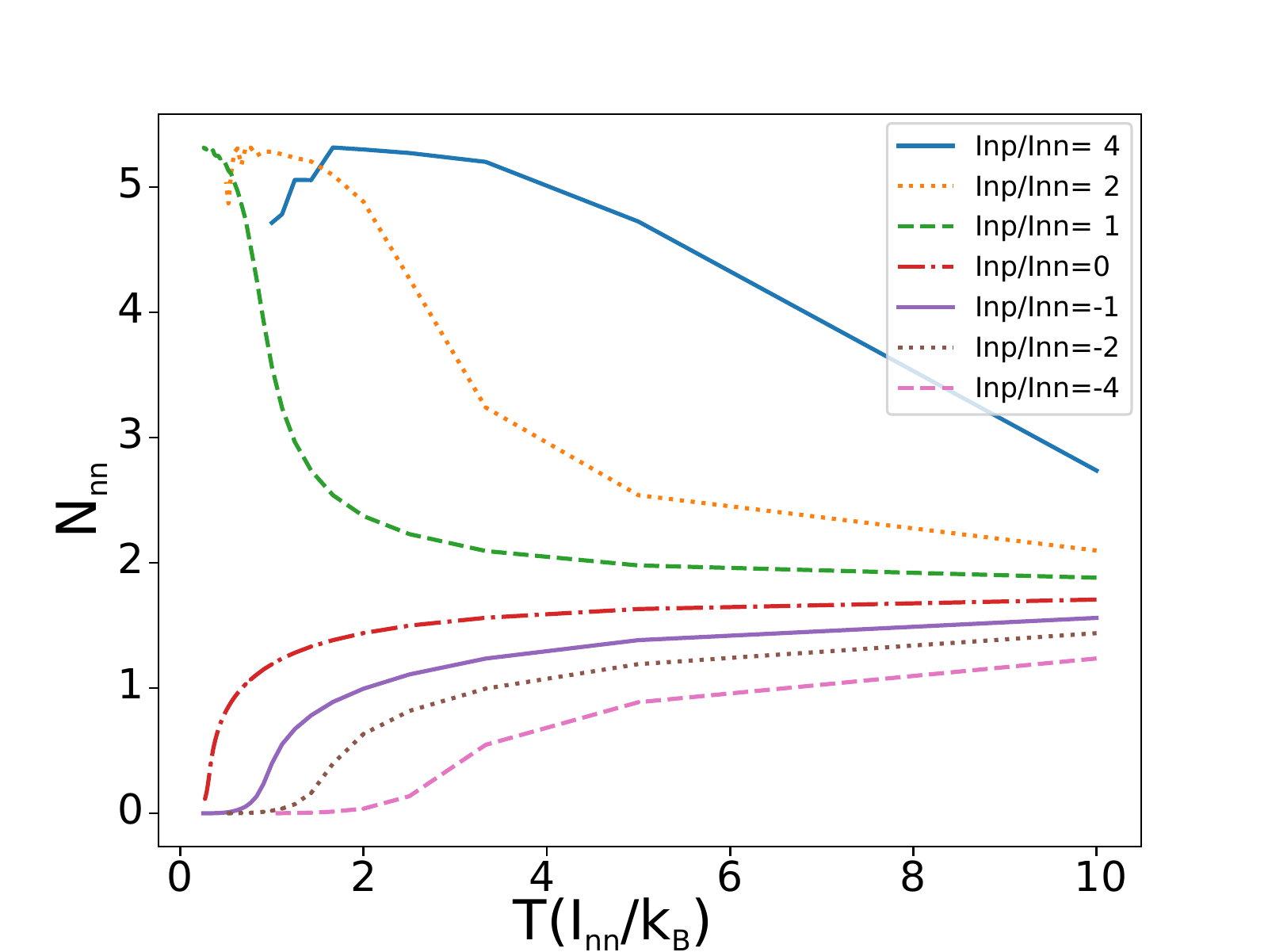}
  \caption{Temperature-induced phase transitions for different values
    of $I_{np}/I_{nn}$. Temperature in units of $I_{nn}/k_B$.}
  \label{fig:temperature}
\end{figure}

In the construction of the phase diagram of the system the interaction
$I_{pp}$ was set to 0 and used as reference. The number of anions,
$N_n$, was set to 0.3 times the number of total nodes, $N$. The
various phases can be distinguished by the number of first neighbour
anions of anions per node ($n_{nn}$). An average value of $n_{nn}$ close to 6
corresponds to a homogeneous phase, and a value of about 0 corresponds to
a hexagonal phase, while a value close to 2 corresponds to a striped
phase or to a random distribution of ions. In the latter case,
$\langle n_{nn}\rangle=zN_n/N$, where $z$ is the coordination number, so $\langle n_{nn}\rangle=1.8$ for $N_n/N=0.3$ on a triangular lattice. The resulting phase diagram
is shown in Fig. \ref{fig:phase_diagram}. As can be seen, three
different phases are found. However, if we restrict ourselves to
physically feasible values in common ILs (with positive $I_{nn}$ and
negative $I_{np}$), only disordered and hexagonal phases are possible
for this fraction of anion-occupied sites. We have investigated the
dependence of the phase on temperature, and the results are
shown in Fig. \ref{fig:temperature}. It can be seen there that, at
high temperatures, the stable phase is the disordered one and at low
temperatures the hexagonal or the homogeneous phase (this latter for
negative values of the attractive parameter).

The main features of the MC phase diagram can be understood by resorting
to a statistical lattice theory. Specifically, it is possible to
describe the patterns observed in the mixture using the well-known
\textit{quasi-chemical approximation},\cite{hill1960,prigogine} given
by
\begin{equation}
  \label{eq:mass_action}
  \frac{N_{np}^2}{\left(zN_n-N_{np}\right)\left(zN_p-N_{np}\right)} 
  = \exp{\left(-2\beta\omega\right)},
\end{equation}
where $N_n$ and $N_p$ are the numbers of anions and cations,
respectively, and $2\omega=-(I_{nn}+I_{pp}-2I_{np})$, i.e., the energy
required to change an $nn$ pair and a $pp$ pair into two $np$ pairs.
Formally, eq. \ref{eq:mass_action} is the usual mass-action law for
chemical reactions, which is the reason for the name
\textit{quasi-chemical approximation}, and goes beyond a purely random
distribution introducing independent pairs.\cite{hill1960} We note
that, in eq. \ref{eq:mass_action}, $zN_n-N_{np} = 2N_{nn}$ and
$zN_p-N_{np} = 2N_{pp}$. At high temperatures, eq. \ref{eq:mass_action}
gives $N_{np} = zN_nN_p/\left(N_n+N_p\right)$, which is the value for
random mixing.\cite{prigogine} On the other hand, at low temperatures
there are two possibilities: if $\omega > 0$, the two species are
fully segregated ($N_{np}=0$); if $\omega<0$, the equilibrium
distribution is that which maximizes the first term in eq.
\ref{eq:mass_action}, with the constraints imposed by the lattice. The
result can be written in terms of the number of particles of the
minority component, $N_m$. For the triangular lattice, a system with
$N_m=N/2$ (where $N=N_n+N_p$) is organized in stripes; with $N_m=N/3$
or $N_m=N/4$ in hexagons, and with $N_m=0$ a homogeneous configuration
is found. For intermediate values the system will adopt mixed
configurations of two of them.

\begin{figure}
  \centering
  \includegraphics[width=8.25cm]{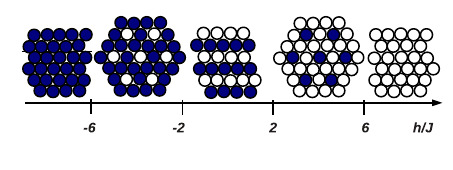}

  \caption{Patterns corresponding to the minimum of the energy per
    lattice site as a function of the ratio between the external
    potential $h$ and $J$, for the particular case of the same
    interaction between the first and the second neighbours, with
    $-I_{np}=I_{pp}=I_{nn}=J/2$ and $\mu_{p}=-\mu_{n}=h$. The
    positively and negatively charged ions are represented as open and
    filled circles respectively. }
  \label{fig:scheme_mu}
\end{figure}

On the other hand, 
the number of anions in the ground state of the mixture in the model
of eq. \ref{eq:total_energy_1}
can be estimated minimizing 
$f=E-\left(\mu_n-\mu_p\right)N_n$, where $E$ is given by eq.
\ref{eq:total_energy_1} and $\mu_n$ and $\mu_p$ are the
electrochemical potentials of the two ionic species. Assuming, for
simplicity, a random distribution of ions
the equilibrium condition $df/dN_n=0$ yields the expression
\begin{equation}
  \label{eq:minimum}
  N_n=\frac{z\left(I_{pp}-I_{np}\right)+\left(\mu_n-\mu_p\right)}{z\left(I_{pp}-2I_{np}+I_{nn}\right)}.
\end{equation}
This result shows that it is possible to achieve any concentration of
anions by tuning the difference between the surface electrochemical
potentials, independently of the temperature. Accordingly, it should
be possible to control the surface pattern by carefully adjusting the
electrochemical potentials, e.g., modifying the electrode's potential.
A scheme of the various possible 2D patterns as a function of the
electrochemical potentials ($h$) is shown in Fig.
\ref{fig:scheme_mu}. It is visible there that the patterns coincide
with our MD and MC predictions.

The variety of systems in which our 
simulations show the presence of striped and hexagonal patterns
indicates that we are probably in front of a general feature of
surface-confined ILs, a question which requires a sound theoretical
foundation. To address this issue, we have used the phenomenological
Landau-Brazovskii (LB) model.\cite{brazovskii:75:0} According to it,
systems with mesoscopic-scale inhomogeneities can be described by the
functional
\begin{align}
 \nonumber {\cal L}_B\left[\phi\right]=\int d{\bf r}&\left[ f\left(\phi\left({\bf
r}\right)\right)+\frac{\beta V_2}{2}|\nabla \phi\left({\bf
  r}\right)|^2+ \right.\\
\label{braz}&\left.\frac{\beta
V_4}{4!}\left(\nabla^2 \phi\left({\bf r}\right)\right)^2\right],
\end{align}
where  $\phi({\bf r})$ is the order parameter (OP). 
Typically, it is assumed that
\begin{equation}
  f(\phi)=(A_2/2+\beta V_0)\phi^2 + A_3\phi^3/3!+A_4\phi^4/4!
  \label{Alina}
\end{equation} with $A_2,A_4>0$. $A_n$ and $V_n$ 
are phenomenological parameters.  At the mean-field (MF) level,  the equilibrium OP corresponds to the minimum of ${\cal L}_B$.
When $V_2<0$, the inhomogeneous structure is favoured by the second term in eq. \ref{braz}, because in the presence of inhomogeneities ($\nabla \phi({\bf r})\neq 0$) this term leads to a decrease of the functional.  When $V_4>0$,
the inhomogeneous structure is disfavoured by the third term in
eq. \ref{braz}. In the case of $V_2<0$ and $V_4>0$,
competition between these two terms leads to a
finite length-scale of inhomogeneities, $2\pi/k_b$, with $k_b^2=-6
V_2/V_4$.\cite{brazovskii:75:0} {The effective field theory given by eq. \ref{Alina} is, in some sense, a generalization of Limmer's theory proposed in ref. \citenum{Limmer-2015}, where the OP represents the local excess charge density, and the Landau functional for a phase-separating mixture is supplemented with the Coulomb interactions between the ionized species. In our LB theory the specific and Coulombic interactions are treated at the same footing, and the neighbourhood of opposite charges is favoured when $V_2<0$ and $V_4>0$. Moreover, we explicitly consider possible asymmetric solutions by taking into account the term $A_3\phi^3/3!$, neglected in ref. \citenum{Limmer-2015}. As we show later, this term plays a crucial role in the symmetry of the ordered structure.}

At high temperature the Brazovskii functional takes the minimum for
$\phi({\bf r})=0$, i.e., for the disordered phase.  Ordered patterns
with spatially periodic $\phi({\bf r})$ can correspond to the
minimum of ${\cal L}_B$ for sufficiently low $T$, since for decreasing
$T$ (increasing $\beta$) the role of the last two terms in
eq. \ref{braz} increases. In 2D systems, the stable ordered pattern
corresponds to parallel stripes for relatively small values of
$|A_3|$, or to a hexagonal arrangement of circular domains with
$\phi({\bf r})>0$ or $\phi({\bf r})<0$ for negative or positive values
of $A_3$ with intermediate magnitude, respectively.  In addition to
the ordered phases, a disordered phase is present for very large
$|A_3|$.  The values of $|A_3|$ at the transitions parallel stripes -
hexagonal structure, and hexagonal structure - disordered phase are
functions of the remaining parameters in eq. \ref{braz}, and both
increase with decreasing $T$.\cite{ciach:10:1}

In the presence of fluctuations, there are some changes on the phase
diagram described above.  The stability region of the disordered phase
enlarges (fluctuations destroy the order), and a striped phase with
orientational and without translational order
appears.\cite{Pekalski-2014,almarza-2014} In this phase, defects destroy the
translational order, but a preferred direction of the stripes remains (anisotropic structure).  At sufficiently low $T$ the
general sequence of the structures for increasing $|A_3|$: oriented
stripes - hexagonal - disordered - is preserved beyond MF.  The
disordered phase is isotropic, but the correlation function exhibits
oscillatory decay at relatively low $T$, signaling short-range
periodic order.

Predictions of the LB theory agree with the patterns that are formed
in thin magnetic films with competing ferromagnetic and dipolar
interactions,\cite{barci:13:0} in amphiphilic
systems~\cite{leibler:80:0,fredrickson:87:0,seul:95:0,gompper:94:3,ciach:01:2}
and in colloidal systems with competing short-range attraction and
long-range repulsion
(SALR).~\cite{imperio:06:0,archer:08:0,ciach:10:1,ciach:13:0,edelmann:16:0,zhuang:16:0} In
magnetic films, the OP is identified with local magnetization.
In the latter two examples, $\phi({\bf r})$ corresponds to the
local deviation from the space-averaged value of the concentration
difference between the polar and apolar components, and to the volume
fraction of particles, respectively.
 
The patterns predicted by the LB theory agree very well with the
charge inhomogeneities observed in the layer near the graphene electrode and with the predictions of our lattice model. The sequence of phases agrees with that in Fig. \ref{fig:scheme_mu}, and the short-range order in the disordered
phase agrees with the upper right snapshot in Fig. \ref{fig:phase_diagram}.
In order to describe these inhomogeneities by means of the functional
in eq. \ref{braz}, the OP $\phi({\bf r})$ must be identified, as has also been previously done by Ebeling \emph{et al.},\cite{Ebeling-2016} with the
deviations of the local charge density,
$\rho_q({\bf r})=e(\rho_p({\bf r})-\rho_n({\bf r}))$ from its
space-averaged value, $\bar\rho_q=e(\bar\rho_p-\bar\rho_n)$, where
$e$ is the elementary charge and $\rho_p({\bf r})$ and
$\rho_n({\bf r})$ denote the local volume fraction of cations and
anions respectively (controllable by different means such as voltage,
concentration of additives, etc.).  In equilibrium, $\bar\rho_q$ is determined by the external conditions.

To apply the continuous theory eq. \ref{braz} to the structure formed at
the microscopic scale, we adopt the concept of the mesoscopic volume
fraction.\cite{ciach:11:0} Let us assume, for clarity, that the ions
and the neutral molecules are roughly spherical and have nearly the
same diameter $D$. We choose this size as the scale of coarse-graining
or smearing of the microscopic density. The mesoscopic volume fraction
of cations at the point ${\bf r}$, $\rho_p({\bf r})$, is equal to the
fraction of the volume of the sphere with the center at ${\bf r}$ and
diameter $D$ that is covered by the cations.
$\rho_p({\bf r})$ decreases continuously from $1$ when the center of
the cation is shifted from ${\bf r}$.
The mesoscopic volume fractions of the anions and the neutral
component are defined in an analogous way. Note that with this
definition each microscopic state is characterized by continuous
fields, so differential operators can be applied.
 
In order to predict which pattern will be stable under given
thermodynamic conditions, we need to know the dependence of $A_3$ on
the measurable quantities.  Here we follow the analysis of ref.
\citenum{ciach:13:0}, where a link between the phenomenological theory
and the statistical thermodynamics was established by a systematic
coarse-graining procedure for the SALR system.  We assume that in our
case ${\cal L}_B$ 
is equal to the excess grand potential associated with lateral
inhomogeneities in the layer near the graphene electrode. Next we assume that
the neutral molecules and/or the ionic vacancies (if present) in the near-surface layer are homogeneously distributed. With this assumption, the sum
of the volume fractions of the anions and the cations in the
considered layer is approximately constant, and ${\cal L}_B$ depends
only on the lateral charge distribution
$\phi({\bf r})=\rho_q({\bf r})-\bar\rho_q$,

\begin{equation} 
  {\cal L}_B[\phi]=\beta\Omega[\bar\rho_q+\phi]-\beta\Omega[\bar\rho_q].
\end{equation}
The excess grand potential consists of the excess internal energy and
the excess entropy contributions.  The excess entropy,
$S(\bar\rho_q+\phi,\rho)-S(\bar\rho_q,\rho)$, can be Taylor expanded
in terms of $\phi$, and within the local-density approximation we
obtain

 \begin{align}
\nonumber{\cal L}_B&\approx \sum_{n=2}^4
\frac{A_n(\rho,\bar\rho_q)}{n!}\int \!\! d{\bf r} \,\phi({\bf r})^n\\
\label{BT}&+\frac{1}{2}\int \!\! d{\bf r}\int \!\! d\Delta {\bf
r}\phi({\bf r}) \beta V_c(\Delta {\bf r}) g(\Delta {\bf r})\phi({\bf
r}+\Delta {\bf
r}),
\end{align}
where the first and the second term on the right hand side represent the entropy and the
internal energy contributions respectively, with $V_c$ and $g$
denoting the interaction potential and the pair distribution function
respectively. The excess energy can be approximated by the LB-type
expression, as shown in ref \citenum{ciach:13:0}.  The parameters
$V_n$ can be expressed in terms of the Coulomb interactions. When $g$
is such that $g(r)=0$ for $r<D$, then $V_2<0,V_4>0$, as in the LB
theory.  We assume that the entropy can be approximated by the entropy
of mixing of the ions and the neutral molecules,
$S/V=-k_BT\sum_{i=p,n,s} x_i\ln x_i$, where $x_i$ is the mole fraction
of the $i$-th component in the considered volume $V$ and the subscript
$s$ is for the neutral component. We also assume that the above form for
the entropy is valid locally (i.e., in each region with the center
at ${\bf r}$ and linear size $D$, with $x_i$ taking the local value
$x_i({\bf r})$). In the simplest case of equal sizes of all the
components, $x_i=\rho_i/\rho_{tot}$, where
$\rho_{tot}=\sum_{i=p,n,s}\rho_i$.  By assumption,
$\rho_s=\rho_{tot}-(\rho_p+\rho_n)$ is independent of the lateral
position.

The parameters $A_n(\rho,\bar\rho_q)$ in eq. \ref{BT} can be easily
obtained. We get $A_2,A_4>0$, and
 \begin{equation}
 \label{A3}
  A_3=\frac{1}{2\rho_{tot}}\Big[
\frac{1}{(\rho-\bar\rho_q)^2}-\frac{1}{(\rho+\bar\rho_q)^2} \Big],
\end{equation} 
which can be positive or negative depending on the sign of
$\bar\rho_q$.  Note that when $|\bar\rho_q|\ll \rho$, then $A_3$ is
very small and stripes are expected by the LB functional for a weakly
charged electrode.  When $\bar\rho_q<0$ (positively polarized
electrode), then $A_3<0$, and the hexagonal pattern of the positive
charges in the negatively charged background occurs for a range of
$|A_3|$ depending on $T$.  For $\bar\rho_q>0$ (a layer near the
cathode) $A_3>0$, and the ``negative'' of the above pattern, i.e.,
circular domains of negative charge forming a hexagonal structure
appear for some range of $A_3$.  The second case (i.e.,
$\bar\rho_q>0$) corresponds to our simulations performed in the
presence of the graphene cathode.

From eq. \ref{A3}, it follows that $|A_3|$ increases when
$|\bar\rho_q|$ increases with fixed $\rho$, or $\rho$ decreases with
fixed $\rho_q$. It means that the transition between oriented stripes
and hexagons can be induced by increasing the charge imbalance in the
adlayer (increasing $|\bar\rho_q|$), and/or by adding a neutral
component (decreasing $\rho$). Indeed, any physical perturbation which
couples to the OP $\phi$ (differences in the charge density) is
candidate to trigger structural transitions in the 2D confined IL
close to the interface. This behaviour agrees with all simulation
results. The change of the pattern from stripes to hexagons
was observed in simulations when $|\bar\rho_q|$ was increasing, and/or
$\rho$ was decreasing. In particular, in Table 1 hexagons are
observed for large charge imbalances ($\eta<0.2$), while the oriented
stripes are associated to small ones ($\eta>0.2$).


{To conclude, it is also noteworthy that these patterns could be also seen as the time-averaged outcome of the 2D dynamics of strongly adsorbed IL mixtures confined in the region close to the electrode, dynamics which has been recently investigated by means of NMR relaxometry.\cite{Kruk-2016} On the other hand,} striped and hexagonal structures have also been reported as Turing patterns driven by a reaction-diffusion model.\cite{walgraef-2002, clerc-2006} However, as we have shown, it seems
that an equilibrium model is enough to explain the observed patterns
at the electrochemical interface.\cite{Pekalski-2014} {The relation of the hereby reported structural patterns with the dynamics of ILs under confinement, as well as} with the Turing-like mechanism and its possible equivalence to the formalism hereby reported, deserve further attention and will be developed elsewhere.

\section{Conclusions}
Our MD simulations show that, in agreement to previously reported works,\cite{Merlet-2014, rotenberg-2015,Ebeling-2016,dudka-2016,Docampo-2016} there are, at least, two different possible patterns (striped and hexagonal) for ILs in the Stern layer formed at IL-graphene interfaces, corresponding to the disorder (homogeneous) and ordered (crystal-like) phases reported in the literature. We have shown that the formation of these 2D structures can be theoretically explained on very general thermodynamic grounds using the Landau-Brazovskii phenomenological theory, thus showing that they are the result of an equilibrium of forces in the adlayer. The fact that these structures can be replicated, in the case of a pure 2D IL, using a simple lattice model
with only nearest-neighbour interactions shows that the most relevant
interactions in the pattern formation in such dense ionic environments
are short-ranged core interactions. Additionally, our results show that the structural patterns are also triggered by changes in the structure of the electrode (vacancies). Hence, we proved that,
independently of its origin, any perturbation affecting
the net balance of electric charge at the IL-graphene interface (described
here by the order parameter ($\phi$(r)) can
trigger transitions between the possible 2D
patterns. Thus, given the potential influence of the lateral structure of the adlayer on the charging dynamics, our findings may open
new ways to induce structural transitions 
at those interfaces, which can be of technological
interest.

Of course, molecular details may have a significant effect on the results at the quantitative level. In particular, 
a more accurate approximation for the entropy in specific systems will lead to some corrections to our expression for $A_3$ in eq. (\ref{A3}).
Thus, the values of the average charge $\bar\rho_q$ and concentration $\rho$ of ions at the transitions between the different structures will be somewhat different. However, our results, as well as others previously reported, indicate that the sequence of structures should remain the same for a large class of systems, suggesting a universal behaviour in some way analogous to the universal sequence of gas, liquid and solid phases that occurs 
in majority of substances, independently of the molecular details. 
These universal features are predicted by the generic model that successfully describes many other systems exhibiting pattern formation, including systems composed of molecules as complex as lipids or block copolymers.\cite{seul:95:0,gompper:94:3} 


Finally, we must mention that although in this work we have focused
entirely on 2D pattern formation in ILs confined between  graphene walls, it is evident, as indicated above, that the formation of these 2D structures greatly influences (and is influenced by) the 3D structure of the electric double layer. This, indeed, has been reported by Ebeling \emph{et al.},\cite{Ebeling-2016} who found a 3D zincblende-type ionic crystal-like organization in the interface between propylammonium nitrate and highly ordered pyrolytic graphene. The theoretical study of these 3D effects would add new insights into the behaviour of these systems as it would the analysis of packing fraction effects on the transitions between ordered and disordered phases in ILs in such confined geometries. Work in this direction is now in progress.


\section{Acknowledgement}
  Prof. R. M. Lynden-Bell is greatly acknowledged for her most useful
  comments, suggestions and careful reading of our manuscript. The
  authors wish to thank the financial support of the Spanish Ministry
  of Economy and Competitiveness (Projects MAT2014-57943-C3-1-P,
  MAT2014-57943-C3-3-P and FIS2012-33126). Moreover, this work was
  funded by the Xunta de Galicia (AGRUP2015/11 and GRC ED431C
  2016/001). All these research projects were partially supported by
  FEDER. H. M.-C. and J. M. O.-M. thank the Spanish Ministry of Education for their FPU grants. Facilities provided by the Galician Supercomputing Centre
  (CESGA) are also acknowledged. Funding from the European Union (COST
  Actions CM1206 and MP1303) is also acknowledged. This project has
  also received funding from the European Union's Horizon 2020
  research and innovation programme under the Marie Sk\l{}odowska-Curie
  grant agreement No 734276, and from the National Science Center grant 2015/19/B/ST3/03122.

\bibliography{bibliography}

\bibliographystyle{rsc} 
\end{document}